\documentclass[twocolumn, times, tighten]{aastex62}
\usepackage{amsmath}

\shortauthors{Krapp et al.}

\begin{document}

\title{\sc{Streaming Instability for Particle-Size Distributions}}

\shortauthors{Krapp et al.}
\shorttitle{Multispecies streaming instability} 
\author{Leonardo Krapp}\affil{Niels Bohr International Academy, Niels Bohr Institute, Blegdamsvej 17, DK-2100 Copenhagen \O{}, Denmark}
\email{krapp@nbi.ku.dk}
\author{Pablo Ben\'itez-Llambay}\affil{Niels Bohr International Academy, Niels Bohr Institute, Blegdamsvej 17, DK-2100 Copenhagen \O{}, Denmark}
\email{pbllambay@nbi.ku.dk} 
\author{Oliver Gressel}\affil{Niels Bohr International Academy, Niels Bohr Institute, Blegdamsvej 17, DK-2100 Copenhagen \O{}, Denmark}\affil{Leibniz-Institut f{\"u}r Astrophysik Potsdam (AIP), An der Sternwarte 16, 14482 Potsdam, Germany}
\author{Martin E. Pessah}\affil{Niels Bohr International Academy, Niels Bohr Institute, Blegdamsvej 17, DK-2100 Copenhagen \O{}, Denmark}

\begin{abstract}
The streaming instability is thought to play a central role in the early stages of planet formation by enabling the efficient bypass of a number of barriers hindering the formation of planetesimals.  
We present the first study exploring the efficiency of the linear streaming instability when a particle-size distribution is considered. We find that, for a given dust-to-gas mass ratio, the multi-species streaming instability grows on timescales much longer than those expected when only one dust species is involved. In particular, distributions that contain close-to-order-unity dust-to-gas mass ratios lead to unstable modes that can grow on timescales comparable, or larger, with those of secular instabilities. We anticipate that processes leading to particle segregation and/or concentration can create favourable conditions for the instability to grow fast.  
Our findings may have important implications for a large number of processes in protoplanetary disks that rely on the streaming instability as usually envisioned for a unique dust species.
Our results suggest that the growth rates of other resonant-drag-instabilities may also decrease considerably when multiple species are considered. 
\end{abstract}


\section{Introduction}

\begin{figure*}[htb!]
	\centering
	\includegraphics[]{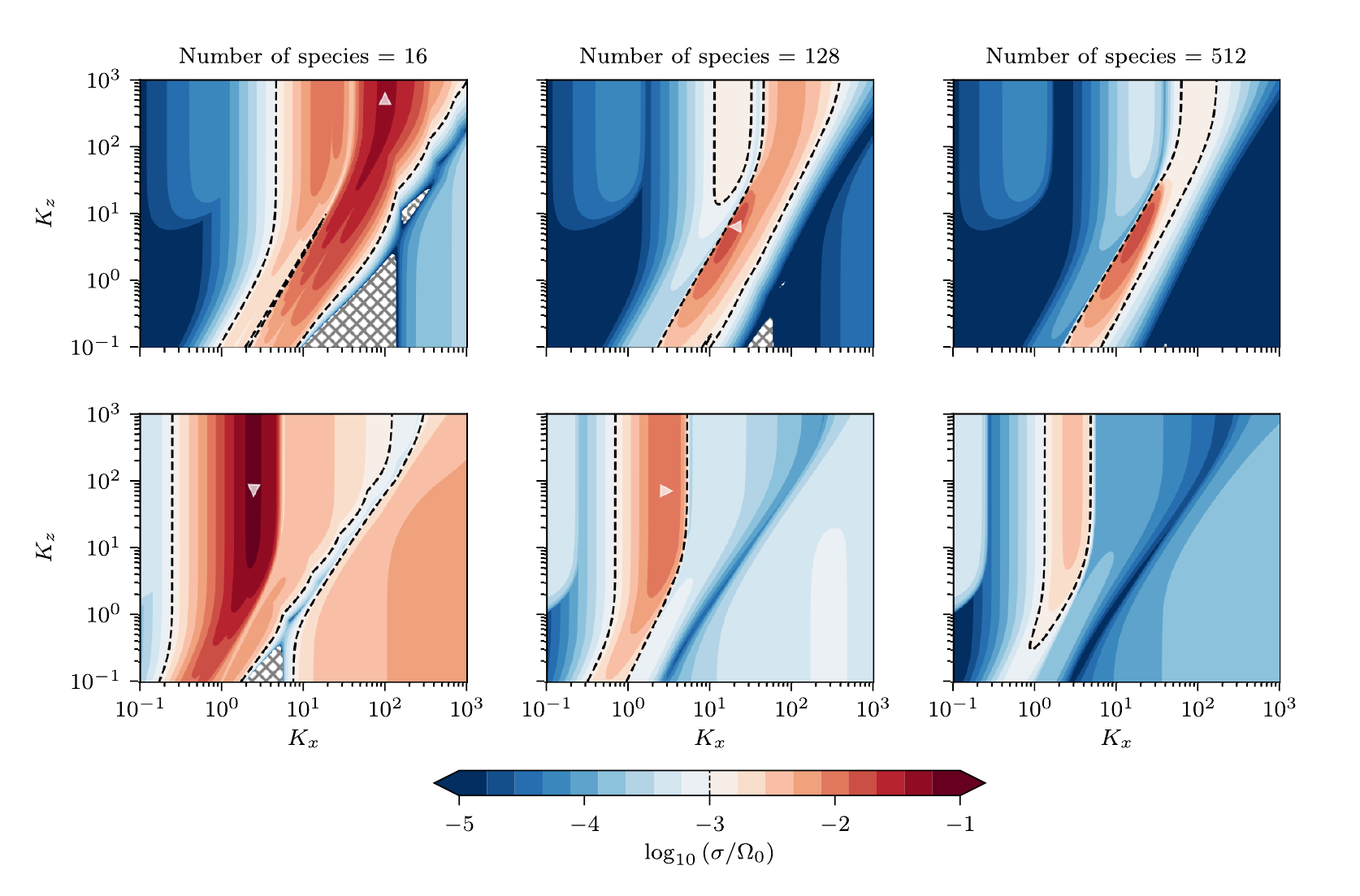}
	\caption{Color map displaying the growth rate $\sigma$ of the most unstable mode for the multi-species streaming instability as a function of the wavenumbers $K_x$ and $K_z$. Results are shown for two number-density distributions with power-law of index $q=-3.5$ in the particle-size/Stokes number and equal dust-to-gas mass ratio, $\epsilon=1$, for Stokes numbers logarithmically spaced in $\Delta T_{\rm s}^{\rm I} = \left[10^{-4}\,,10^{-1}\right]$ (upper panels) and $\Delta T_{\rm s}^{\rm II} = \left[10^{-4}\,,1\right]$ (lower panels) and for an increasing number of dust species $N=16, 128, 512$. Dashed lines denoting $\sigma = 10^{-3}\Omega_0$, mark the turning point of the divergent color palette. Hatched regions are stable. White triangles correspond to the fastest growing modes whose temporal evolution we checked independently using the code FARGO3D (see Fig.\,\ref{fig:convergence}).
}	
\label{fig:maps_vs_nbin}
\end{figure*}

The building blocks of planetary bodies are kilometer-sized planetesimals, which are believed to form and grow from small dust-particles present in the protoplanetary disk. 
When growing from micro- to meter-sized objects, the particles need to overcome the so-called ``growth barriers'' \citep[see][for a review]{testi2014}. In particular, under typical conditions expected in protoplanetary disks, the radial-drift barrier prevents particles from growing beyond centimeter scales.  This corresponds roughly to the particle size for which the timescale involved in particle growth is comparable to the timescale associated with their radial drift \citep{Whipple1972, Weidenschilling1977, Nakagawa1986}. One mechanism envisioned to overcome this barrier is the streaming instability \citep{Youdin2005, Youdin2007, Jacquet2011, Auffinger2018}.

Even though there have already been a handful of papers reporting numerical simulations exploring the effects of dust-size distributions in the non-linear outcome of the streaming instability \citep[see e.g.][]{Bai2010b, Schaffer2018}, a systematic study addressing its linear regime is yet to be presented.
Such a study is necessary for a number of reasons. Conceptually, it provides a more sensible framework to assess the efficiency of the streaming instability in more realistic models of protoplanetary disks. From a computational perspective, it provides valuable benchmarks against which to test numerical codes \citep[e.g.][]{Benitez-Llambay2019}, as well as an idea of the numerical requirements to recover the proper evolution of the instability during its early (linear) phase.

In this Letter, we present the first study of the linear phase of the streaming instability involving a distribution of dust-particles of different sizes. Our systematic exploration of parameter space allows us to provide the growth rate of the most unstable mode as a function of the dust-to-gas mass ratio, particle-size range, and number of dust species considered for describing a particle-size distribution. 

\section{Multi-species Streaming Instability}
\label{sec:method}

\subsection{Dust-size Distribution}
\label{subsec:dust_distribution}
We consider an underlying (continuous) number-density distribution of particles, which is a power-law of index $q$ in the particle size, $a$
\citep{Dohnanyi1969,Mathis1977}. In what follows, we take $q=-3.5$ for definitiveness\footnote{We have obtained similar results to those presented here using $q=-4$, which corresponds to a distribution with equal mass per logarithmic bin in Stokes number. Our analysis can be easily generalized to accommodate for more sophisticated distributions.} \citep[see e.g.,][]{Garaud2004}. We consider the Epstein regime in which the Stokes number is proportional to the particle size, i.e., $T_{\rm s} \propto a$.

A discrete approximation of this dust-size distribution is characterized by the total gas-to-dust mass ratio, $\epsilon$, a range of Stokes numbers properly bound by a minimum and a maximum, $\Delta T_{\rm s} = [T_{\rm s, min}, T_{\rm s, max}]$, and the total number of species $N$ associated with a distinct Stokes number,  $T_{{\rm s},i}$, with $i=1, \ldots, N$. The total mass of the distribution remains constant when varying $N$,  i.e., $\sum_{i=1}^N \epsilon_i = \epsilon$, where $\epsilon_i$ is the dust-to-gas mass ratio associated with a given dust species.
We define the Stokes numbers characterizing the distribution evenly in logarithmic scale\footnote{The results of this work do not change qualitatively if we distribute the Stokes numbers so that they all contain the same dust-to-gas mass ratio per species.}.

A robust discrete approximation of the underlying particle-size distribution should in principle lead to a dynamical model that converges as the number of dust species considered increases, i.e., as the continuous limit is approached. Therefore, it is of particular interest to understand the sensitivity of the results obtained with respect to the number of dust species, $N$, used to describe the underlying dust-size distribution.

\begin{figure*}
	\centering
	\includegraphics[]{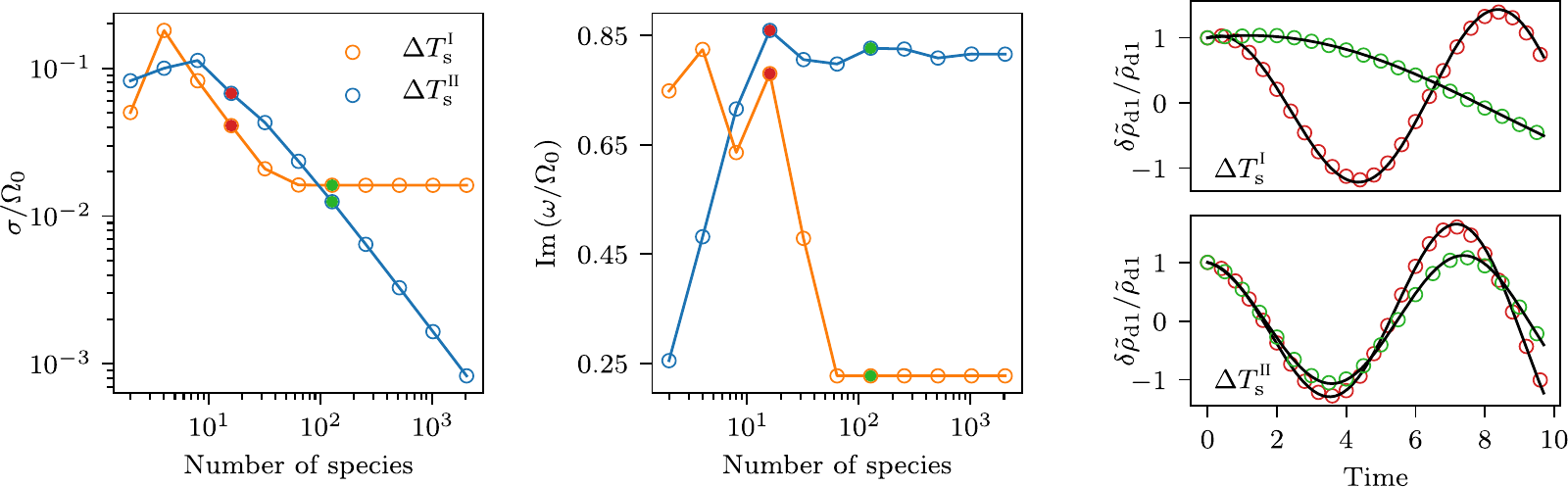}
	\caption{Real (left panel) and imaginary (middle panel) parts for the eigenvalues corresponding to the most unstable modes for the particle distributions with dust-to-gas mass ratio $\epsilon=1$ and Stokes numbers in the intervals $\Delta T_{\rm s}^{\rm I} = \left[10^{-4}\,,10^{-1}\right]$ (orange) and $\Delta T_{\rm s}^{\rm II} = \left[10^{-4}\,,1\right]$ (blue) as a function of particle species number $N$. The two rightmost panels show the time evolution of the density fluctuation of dust species 1, $\delta\rho_{1}$, for the most unstable eigenmode.  Red and green unfilled circles show the solutions obtained with FARGO3D for 16 and 128 dust-species, respectively, using $32$ cells per wavelength. The solid black lines correspond to the solutions of the linear mode analysis described in Section \ref{subsec:linear_modes}.}
	\label{fig:convergence}
\end{figure*}

\medskip
\subsection{Linear Modes in Fourier Space}
\label{subsec:linear_modes}
The equations describing the dynamics of a gas coupled to $N$ dust species via drag forces in the framework of the shearing box, together with the analytical steady-state background solution, have been recently derived in \citet{Benitez-Llambay2019}, see Section 3.5.
Linear axisymmetric perturbations with respect to the steady-state background lead to $4(N+1)$ (continuity and momentum) equations describing the multiple-species streaming instability, presented in their Appendix E.
A given eigenmode of this linear system evolves in space and time according to $\textrm{Re}[\delta \hat{f}(k_x, k_z) \, e^{i(k_x x + k_z z) - \omega t}]$. Here,  $\delta \hat{f}(k_x, k_z)$ are the $4(N+1)$-dimensional  (complex) eigenvectors in Fourier space, spanned by the wavenumbers $(k_x,k_z)$, associated with the (complex) eigenvalue $\omega(k_x,k_z)$. In the context of the streaming instability, it is customary to work with dimensionless wavenumbers $K = H_0^2 k/R_0$, where $H_0$ is the disk sale-height at the fiducial radius $R_0$, where the shearing box is centered, and to use the Keplerian angular frequency, $\Omega_0 \equiv \Omega_{\rm K}(R_0)$, to scale the eigenvalues. 

The early evolution of the instability is governed by the unstable modes -- i.e., those with $\textrm{Re}(\omega) <0$ -- with maximum growth rate\footnote{i.e., $\sigma = \textrm{max} \{|\textrm{Re}(\omega)| : \textrm{Re}(\omega) < 0\}$.} $\sigma$.  In order to identify these modes, we consider the space spanned by the set $(K_x,K_z)$ when each normalized wavenumber takes values in the range $[10^{-1},10^3]$. Our fiducial grid is evenly spaced in logarithmic scale and contains $260$ cells in each direction. Given a dust-size distribution, the dynamical evolution of a specific mode is completely determined by the spectrum of $4(N+1)$ complex eigenvalues $\omega$. This spectrum is found using the function  \verb|eig| of NumPy \citep{Walt2011}, which uses LAPACK routines for complex nonsymmetric matrices \citep{lapack}. 

\subsection{Fastest Growing Modes \/-- Two Test Cases}
\label{sec:map}
We consider two discrete particle-size distributions both with $\epsilon = 1$, but spanning two overlapping ranges of Stokes numbers:  $\Delta T_{\rm s}^{\rm I} = \left[10^{-4}\,,10^{-1}\right]$ and $\Delta T_{\rm s}^{\rm II} = \left[10^{-4}\,,1\right]$. We compute the growth rate using $N \in \{16,128,512\}$ dust species.  These considerations lead to six different eigenvalue problems that are solved to find the fastest growing modes as a function of  $(K_x, K_z)$. The results corresponding to $\Delta T_{\rm s}^{\rm I}$ and $\Delta T_{\rm s}^{\rm II}$,  for each of the adopted $N$-values, are shown in the upper and lower panels of Fig.\,\ref{fig:maps_vs_nbin}, respectively.  
For  $\Delta T_{\rm s}^{\rm I}$, the upper panels show a maximum growth rate that converges with increasing dust species to $\sigma \simeq  1.6\times10^{-2}\Omega_0$ (see Fig.\,\ref{fig:convergence}).  For this distribution, the set of modes that grow fastest converge to a confined region close to the center of the explored domain in $(K_x, K_z)$. In contrast, for $\Delta T_{\rm s}^{\rm II}$ the maximum growth rate decreases monotonically from  $\sigma \simeq 6.7 \times10^{-2} \Omega_0$ for 16 species to  $\sigma \simeq 0.33 \times10^{-2}\Omega_0$ for 512 species (see also Fig.\,\ref{fig:convergence}). 

The sensitivity of the results obtained for the fastest growth rate with respect to the number of species $N$ can be better appreciated in the leftmost panel in Fig.\,\ref{fig:convergence}, which shows the growth rates of the most unstable modes for the two distributions with $\Delta T_{\rm s}^{\rm I}$ (orange line) and $\Delta T_{\rm s}^{\rm II}$ (blue line), when the number of dust species doubles from $N=2$ to $N=2048$.  For the case $\Delta T_{\rm s}^{\rm I}$ the maximum growth rate converges when using $64$ dust species. This is not the case for $\Delta T_{\rm s}^{\rm II}$, for which the maximum growth rate decreases below $\sigma < 10^{-3} \Omega_0$. However,  in a region around $K_x = 10^{-1}$, the growth rate converges to $\sim 5 \times 10^{-4} \Omega_0$ for $N\ge 64$. This value sets the time scale of the linear instability for $N>2048$ species.

\begin{figure}[]
	\centering
	\includegraphics[]{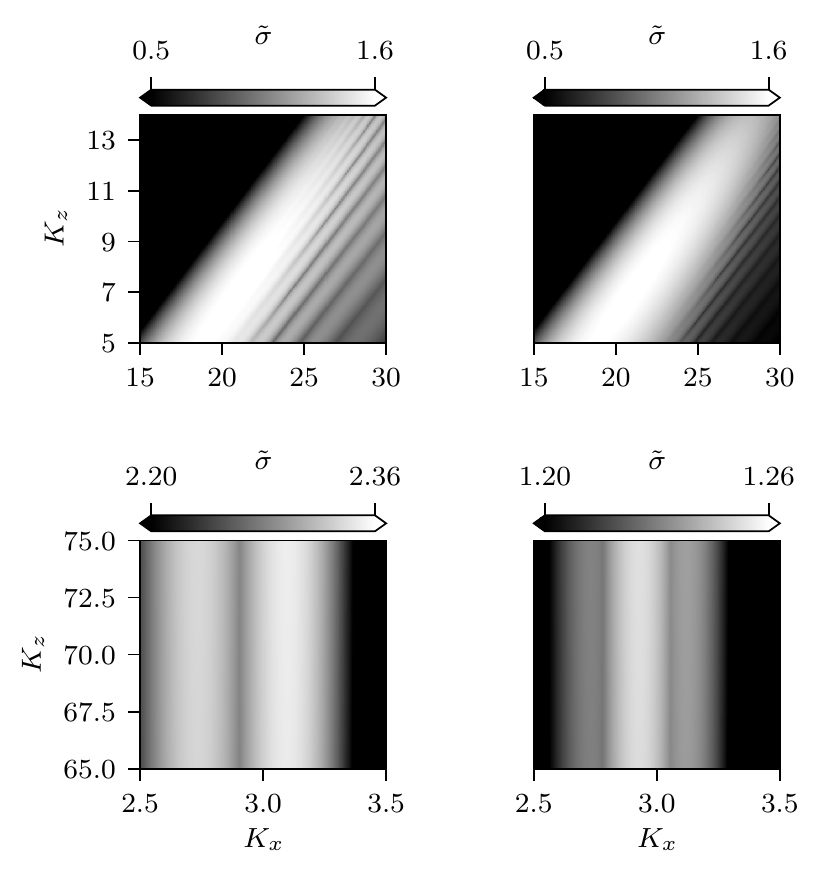}
	\caption{
		High-resolution maps of the normalized growth rate  $\tilde{\sigma} = 10^2\sigma/\Omega_0$. Zoom-in domains surrounding the fastest growing modes for the distributions with Stokes numbers in $\Delta T_{\rm s}^{\rm I} = \left[10^{-4}\,,10^{-1}\right]$ and $\Delta T_{\rm s}^{\rm II} = \left[10^{-4}\,,1\right]$ (upper and lower panels, respectively) for $N= 64$ and 128 species (left and right panels, respectively).}
	\label{fig:zoom}
\end{figure}

In order to shed some light on the strikingly different behaviour exhibited by $\Delta T^{\rm I}$ and $\Delta T^{\rm II}$, we show in Fig.\,\ref{fig:zoom} high resolution maps in $(K_x,K_z)$ zooming-in the neighborhood of the fastest growing modes.  The left and right panels show the maximum growth rate for 64 and 128 species, respectively.  The upper and lower panels correspond to $\Delta T_{\rm s}^{\rm I}$ and $\Delta T_{\rm s}^{\rm II}$ (orange and blue curves in Fig.\,\ref{fig:convergence}), respectively. These maps reveal a pattern with fringes whose number increases linearly with $N$, as they split unstable regions whose growth rates decay also linearly with $N$. When these fringes merge, i.e., their separation is smaller than their width, the growth rates converge. 
It is worth stressing that the imaginary parts corresponding to the most unstable eigenvalues  (shown in the middle panel of Fig.\,\ref{fig:convergence}) do converge as $N$ increases in both cases.
 
The decay observed in the growth rate of the most unstable modes when $N$ increases in the case  $\Delta T_{\rm s}^{\rm II}$ is in stark contrast with the behavior observed for $\Delta T_{\rm s}^{\rm I}$. This warrants a systematic exploration of parameter space including also the dust-to-gas mass ratio. Before embarking on this, and given the complexity of the equations involved in the linear mode analysis describing the multi-species streaming instability, we provide an independent check of our solutions below. 
\begin{figure*}
	\centering
	\includegraphics[width=\textwidth]{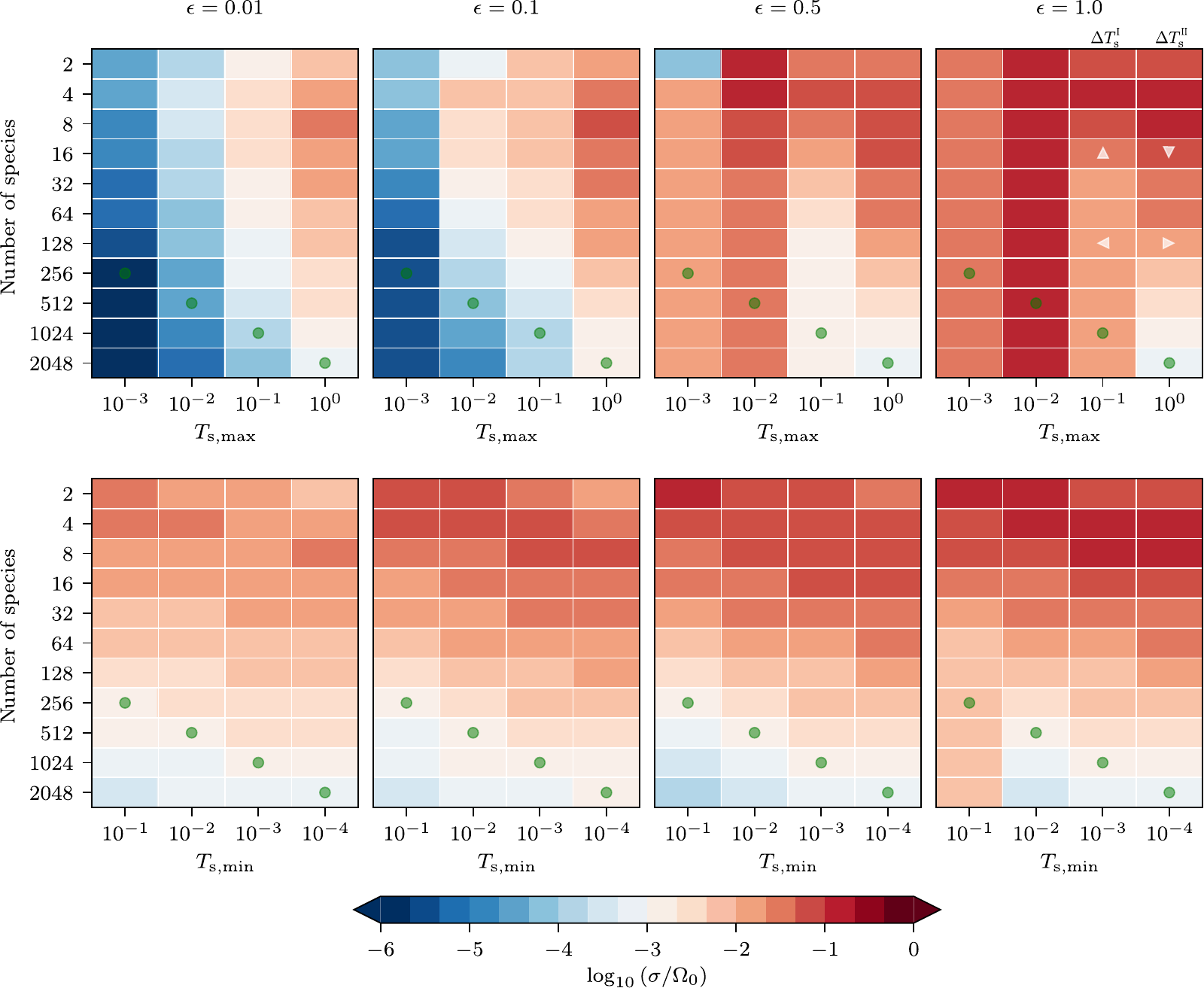}
	\caption{Maximum growth rate for the multi-species streaming instability as a function of the number of the dust-species $N$ for different ranges of Stokes numbers and (fixed) dust-to-gas mass ratio, from left to right each block corresponds to $\epsilon=\{0.01,0.1,0.5,1\}$. The upper panels show the results when considering a fixed $T_{\rm s, min} = 10^{-4}$ and varying $T_{\rm s, max}$. The lower panels show the results when considering a fixed $T_{\rm s, max} = 1$ and varying $T_{\rm s, min}$. The green circles, in each panel, show examples of distributions that have the same number of species per decade in Stokes number. White triangles are used to indicate the correspondence with the distributions used to compute the growth rate maps in  Fig.\,\ref{fig:maps_vs_nbin}.
	}
		\label{fig:results}
\end{figure*}
\subsection{Verification of the Linear Mode Analysis}
\label{subsec:verification}
We test the time evolution of the most unstable modes using the publicly available multifluid code FARGO3D \citep{Benitez-Llambay2016,Benitez-Llambay2019}, following the procedure described in section 3.5.4 of \cite{Benitez-Llambay2019}. 

We consider four representative cases from Fig.\,\ref{fig:convergence} (red and green filled circles) and use the corresponding eigenvectors to initialize four numerical simulations.  The rightmost panels of Fig.\,\ref{fig:convergence} show the time evolution for one of the components (the dust density for species 1, $\delta\rho_1$) for each of these four modes. The solutions obtained with FARGO3D are shown with red and green unfilled circles for 16 and 128 dust-species, respectively. The black solid lines are the solutions obtained from our linear mode analysis described in Section \ref{subsec:linear_modes}.
The excellent agreement between the time evolution of the selected eigenmodes provides additional support to our linear calculations. 
This critically reduces the possibility of potential issues in several steps of our analysis including the derivation of the background equilibrium, the linearization of the perturbed system, and the method used to find the eigenvalues and eigenvectors.
\subsection{Systematic Parameter Space Exploration}
\label{sec:par_explore}
We seek the growth rate of the most unstable mode given a particle-size distribution characterized by the dust-to-gas mass ratio, $\epsilon$, a range of Stokes numbers, $\Delta T_{\rm s} = [T_{\rm s, min}, T_{\rm s, max}]$, and the total number of species $N$. 
We consider four different mass ratios, $\epsilon = \{0.01,0.1,0.5,1\}$ and two sets of intervals in Stokes numbers for which either the minimum is fixed and the maximum varies, i.e.,  $\Delta T_{{\rm s, min}} =  [10^{-4},  T_{{\rm s, max}}]$
with 
$T_{{\rm s, max}} \in  \{10^{-3},10^{-2},10^{-1}, 1\}$,
or 
the maximum is fixed and the minimum varies, i.e., 
$\Delta T_{\rm s,max}=  [T_{\rm s, min}, 1]$
with 
$T_{\rm s, min} = \{10^{-1}, 10^{-2}, 10^{-3}, 10^{-4}\}$.
For each of these intervals in Stokes numbers, we consider an increasing number of dust-species by doubling $N$ from $2$ to $2048$ while keeping the dust-to-gas mass ratio characterizing the distribution constant. This procedure leads to $4\times(4\times2-1)\times11= 308$ independent discrete distributions.

\subsection{Results}
The maximum growth rates for each of the distributions defined above are shown in Fig.\,\ref{fig:results}.
Each panel corresponds to a different dust-to-gas mass ratio $\epsilon$. The rows and columns correspond to a given $N$ and $\Delta T_{\rm s}$, respectively.
Each cell is color-coded according to the logarithm of the maximum growth rate obtained in $(K_x,K_z)$-space, following the method described in Section \ref{sec:method}.

The most relevant outcomes are 
(i) the growth rate of the most unstable modes corresponding to the majority of the distributions with low dust-to-gas mass ratios $\epsilon \lesssim 0.1$ have not converged and decreases below 
$10^{-3}\Omega_0$, independently of $T_{\rm s,max}$. In particular, when $T_{\rm s,min}=10^{-4}$ is fixed, (Fig.\,\ref{fig:results}, upper panels), the upper bound for the grow rate decreases from $10^{-3}\Omega_0$ to $10^{-5}\Omega_0$ as $T_{\rm s, max}$ decreases. (ii) The range of Stokes numbers for which convergence of the growth rate with the number of species is reached increases with $\epsilon$ when $T_{\rm s,min} = 10^{-4}$. (iii) When fixing $T_{\rm s, max}=1$ (Fig.\,\ref{fig:results}, lower panels) convergence of the growth rate, with $N =2048$, is achieved for none of the cases considered but one. The only exception is the case that corresponds to $\epsilon = 1$ and $T_{\rm s, min}=10^{-1}$, for which the most unstable mode has a growth rate $\sigma \simeq 6\times10^{-3}\Omega_0$. In all other cases, the growth rate decreases below $10^{-3} \Omega_0$, independently of the dust-to-gas mass ratio, $\epsilon$. 
\begin{figure*}[]
	\centering
	\includegraphics[]{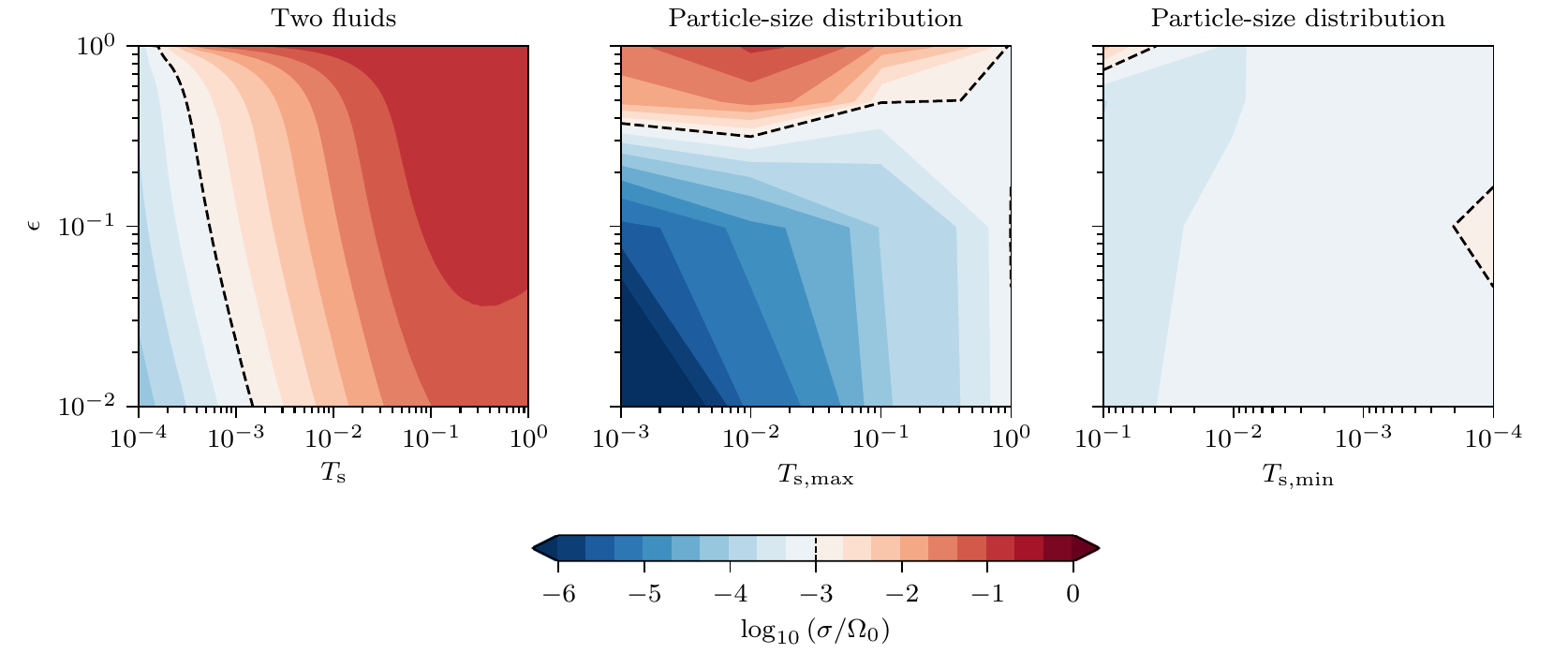}
	\caption{The left panel shows the maximum growth rate corresponding to the classical streaming instability involving only one dust-species. The center and rightmost panel show the maximum growth rate obtained for distributions with $2048$ species when fixing $T_{\rm s, min}=10^{-4}$ and $T_{\rm s,max}=1$, respectively. The dashed line corresponds to $\sigma = 10^{-3}\Omega_0$ in all panels.}
	\label{fig:compare}
\end{figure*}
\begin{figure}[b!]
	\centering
	\includegraphics[]{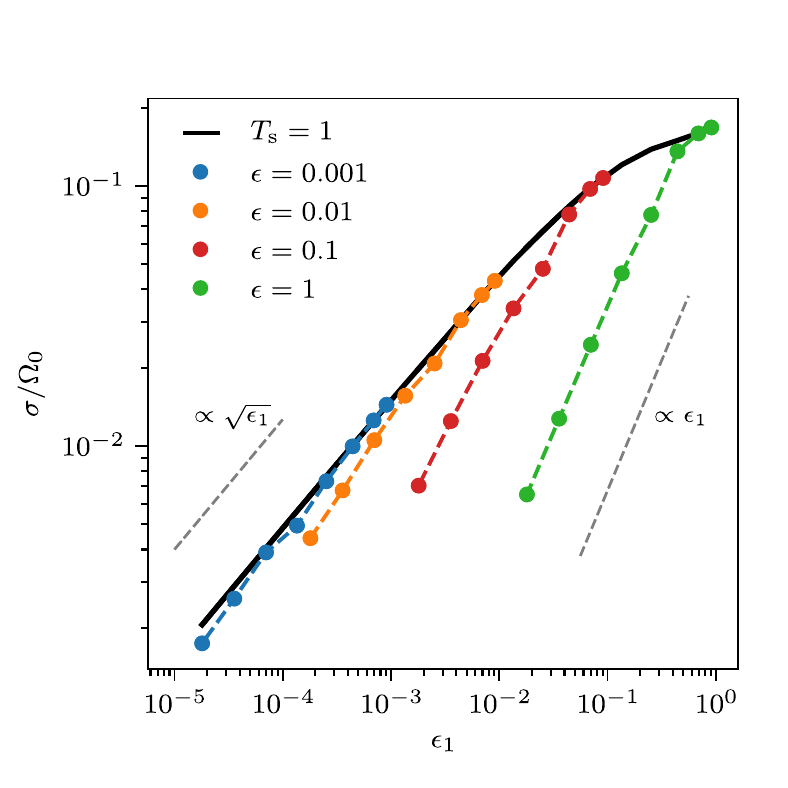}
\caption{Maximum growth rate as a function of the dust-to-gas mass ratio 
	$\epsilon_1\sim\epsilon/N$ of the species with Stokes number that leads to fastest growth when considered in isolation, $T_{\rm s}=1$ in this case (solid black curve). The dashed colored curves correspond to distributions with different dust-to-gas mass ratios (each filled circle is obtained, from right to left, by doubling the number of species $N$ from 2 to 256).  
	}	\label{fig:fig6}
\end{figure}

\section{Outcome and Implications}

We have provided the first systematic study of the linear growth of the multi-species streaming instability. We found two different types of behaviors. On the one hand, there are distributions for which convergence of the growth rates is reached by considering between a handful and a couple of hundred dust species. In the majority of cases we considered, however, we were only able to find upper limits to the growth rates, which are, in many cases, well below the values obtained when only one dust species is involved. This result is better appreciated in Fig.\,\ref{fig:compare}, where we show the maximum growth rates for the classical (gas and one dust-species) streaming instability (leftmost panel) together with those obtained for the distributions with $2048$ species studied in Section \ref{sec:par_explore} (center and rightmost panels). 

Below, we briefly discuss some of the most relevant aspects and consequences of our findings.

\paragraph{Growth Rate Decay and Connection with Resonant Drag Instabilities}
	For a large fraction of the particle distributions we studied, the maximum growth rate $\sigma$ decreases as the number of species $N$ increases (and the mass per species decreases). A concrete example is illustrated in Fig.\,\ref{fig:convergence}, where the growth for the distribution with $\epsilon=1$ and $\Delta T_{\rm s}^{\rm II} = \left[10^{-4}\,,1\right]$ decreases as $1/N$ for sufficiently large $N$. 
	It is natural to compare this to the growth obtained when considering a single dust species with a dust-to-gas mass ratio $\epsilon_1\sim\epsilon/N$ corresponding to the species with Stokes number that in isolation leads to fastest growth. 
	
We found that for $\epsilon < 10^{-2}$ the maximum growth rate of the distribution is in good agreement with the value obtained when considering in isolation the species with largest Stokes number, $T_{\rm s} = 1$ in this case (see Fig.\,\ref{fig:fig6}). However, as the total mass of the distribution considered is increased, a significant difference between these growth rates exists. This is due to two effects that are difficult to disentangle; {\it i}) the background drift-velocity for each individual dust-species is modified because it is sensitive to the total mass of the particle-size distribution and not just the mass per bin (see \citealt{Benitez-Llambay2019}) and {\it ii}) as the ensuing gas perturbation increases, the coupling between species increases and interference among them may not be negligible. 
	
In the limit of very small dust-to-gas mass ratio, $\epsilon\ll1$, we have checked that the $N$ dust-species streaming instability can be well described as the superposition of $N$ different two-fluid instabilities occurring in a seemingly independent way. This suggests that, in this regime, the resonant drag instability (RDI) framework \citep{Squire2018} can help provide insight into the behavior of the multi-species streaming instability.  At sufficiently low dust-to-gas mass ratios the maximum growth rate is expected to decay as $\sqrt{\epsilon_1}\sim\sqrt{\epsilon/N}$. This is indeed obtained for distributions with dust-to-gas mass ratios smaller than $\epsilon=0.01$ (see Fig.\,\ref{fig:fig6}). 

\bigskip
\bigskip
\bigskip
\paragraph{Single-dust Species Models of Streaming Instability}
We report the emergence of an unstable region for $K_x\gg1$ and $K_z \geq 0$, which has not been observed before \citep[see][]{Youdin2005, Youdin2007}. A good example is provided by the two distributions studied in Fig.\,\ref{fig:maps_vs_nbin}. While a comprehensive study of this new unstable region is beyond the scope of this work, we found that the necessary condition for it to appear is the presence of at least two dust-species with opposite background drift-velocities. This observation suggests that it may not be possible to capture the full dynamics of multi-species streaming instabilities using single dust-species models \citep[e.g.,][]{Laibe2014, Lin2017}.

\paragraph{Simulations of Multi-Species Streaming Instability} The successful recovery of known solutions is a key benchmark for any numerical code. Previous numerical studies of the non-linear evolution of the streaming instability with one dust species have been reported to recover, for example, its linear phase  \citep[see e.g.][]{Johansen2007,Balsara2009,Kowalik2013a,Chen2018,Riols2018,Benitez-Llambay2019}.  Even though the linear results have been derived using an Eulerian formalism, it has been shown that numerical codes evolving Lagrangian particles agree very well during the early phases of the streaming instability \citep{Youdin2007}. This suggests that our findings will also hold in the Lagrangian framework. It is possible that a significant decay of the growth rate has not yet been observed in multiple dust-species simulations because of the relatively low number of species that have been used so far \citep[see e.g.][]{Bai2010, Schaffer2018}.

\paragraph{Critical Dust-to-gas Mass Ratio}
Our study suggests that particle-size distributions with $\epsilon \geq 0.5$ are required to allow the multi-species streaming-instability to grow on timescales shorter than $10^{3}\Omega_0^{-1}$ (see Fig.\,\ref{fig:compare}). This, however, depends on the range of Stokes numbers defining the distribution. For example, if $T_{\rm s, max}=1$ the growth rates are smaller than $10^{-3}\Omega_0$, even for large dust-to-gas mass ratios (i.e., $\epsilon=1$). When $\epsilon = 0.5$ and $T_{\rm s, min}=10^{-4}$ the maximum growth rate converges to $\sigma \simeq 10^{-3} \Omega_0$ for $T_{\rm s, max} = 10^{-1}$, and the instability can grow faster for particle-size distributions with $T_{\rm s, max}\leq10^{-2}$. It can also grow faster if the dust-to-gas mass ratio increases to $\epsilon \simeq 1$ for those ranges of Stokes numbers. 
We additionally found that, if the total mass of the distribution decreases below $\epsilon = 0.5 $, the instability develops on timescales of the order of $10^{5}\Omega_0^{-1}$, or even longer, depending on the range of Stokes numbers spanned by the particle distribution.
\paragraph{Planetesimal Formation} We anticipate that the multi-species streaming instability could still be an efficient mechanism to enable planetesimal formation if dust-particles are filtered/segregated according to their size and accumulated somewhere in the disk. This will naturally produce regions with large concentrations of dust with distributions characterized by specific particle-sizes. For instance, vertical sedimentation affected by the presence of winds \citep[e.g][]{Riols2018} or turbulence sustained by the vertical shear instability \citep[e.g.,][]{Lin2019} can favor the accumulation of larger grains at the mid-plane of protoplanetary disks. Some other potential mechanisms for such filtering/segregation are vortices \citep[e.g.,][]{Barge1995,Raettig2015,Ragusa2017}, zonal flows \citep[e.g.,][]{Johansen2009,Dittrich2013,Bethune2016,Krapp2018}, planet-induced pressure bumps \citep[e.g.][]{Zhu2012,Pinilla2012,Weber2018}, and planetary torques \citep{Benitez-Llambay2018,Chen2018}.
Another candidate in this regard is the growth (via dust coagulation) to sizes that are limited by particle-drift, fragmentation or bouncing \citep{testi2014}. Such mechanisms naturally appear to favour accumulating significant amount of mass into (near-)monodisperse populations of particles.

\bigskip
We conclude that a properly resolved particle-size distribution can significantly affect the linear phase of the streaming instability. Depending on the dust-size distribution and dust-to-gas mass ratio, the multi-species instability may only grow on timescales much larger than those expected from the classical 
(gas and one dust-species) case when approaching the continuum limit. In particular, distributions that contain moderate to high dust-to-gas mass ratios (i.e., $\epsilon  \lesssim 1$) will only grow on timescales comparable with those of secular instabilities.
Taken at face value, our results imply that the scope of the streaming instability may be narrowed down profoundly. Nevertheless, processes leading to particle segregation and/or concentration may create favourable conditions for the instability to develop.
Because the growth rate of the multi-species streaming instability depends sensitively on the number of dust species used to represent a distribution, our results may also have important implications for the wider class of resonant-drag-instabilities discussed in \cite{Squire2018}.

\acknowledgements

We thank the referee, Jeremy Goodman, for encouraging us to connect our findings, and their implications, with the more general class of resonant-drag-instabilities.
We thank Troels Haugb{\o}lle for useful discussions and helpful suggestions.
The research leading to these results has received funding
from the European Research Council under the European Union's Horizon 2020 research and innovation programme (grant agreement No 638596) (LK, OG). This project has received funding from the European Union's Horizon 2020 research and innovation programme under grant agreement No 748544 (PBLL).  MEP gratefully acknowledges support from the Independent Research Fund Denmark (DFF) via grant no.\,DFF 8021-00400B. 

\software{ NumPy \citep{Walt2011}, Matplotlib \citep{Hunter2007}}


\begin{thebibliography}{}
	\expandafter\ifx\csname natexlab\endcsname\relax\def\natexlab#1{#1}\fi
	
	\bibitem[{Anderson {et~al.}(1999)Anderson, Bai, Bischof, Blackford, Demmel,
		Dongarra, Du~Croz, Greenbaum, Hammarling, McKenney, \& Sorensen}]{lapack}
	Anderson, E., Bai, Z., Bischof, C., {et~al.} 1999, {LAPACK} Users' Guide, 3rd
	edn. (Philadelphia, PA: Society for Industrial and Applied Mathematics)
	
	\bibitem[{{Auffinger} \& {Laibe}(2018)}]{Auffinger2018}
	{Auffinger}, J., \& {Laibe}, G. 2018, \mnras, 473, 796
	
	\bibitem[{{Bai} \& {Stone}(2010{\natexlab{a}})}]{Bai2010b}
	{Bai}, X.-N., \& {Stone}, J.~M. 2010{\natexlab{a}}, \apj, 722, 1437
	
	\bibitem[{{Bai} \& {Stone}(2010{\natexlab{b}})}]{Bai2010}
	---. 2010{\natexlab{b}}, \apjs, 190, 297
	
	\bibitem[{{Balsara} {et~al.}(2009){Balsara}, {Tilley}, {Rettig}, \&
		{Brittain}}]{Balsara2009}
	{Balsara}, D.~S., {Tilley}, D.~A., {Rettig}, T., \& {Brittain}, S.~D. 2009,
	\mnras, 397, 24
	
	\bibitem[{{Barge} \& {Sommeria}(1995)}]{Barge1995}
	{Barge}, P., \& {Sommeria}, J. 1995, \aap, 295, L1
	
	\bibitem[{{Ben{\'{\i}}tez-Llambay} {et~al.}(2019){Ben{\'{\i}}tez-Llambay},
		{Krapp}, \& {Pessah}}]{Benitez-Llambay2019}
	{Ben{\'{\i}}tez-Llambay}, P., {Krapp}, L., \& {Pessah}, M.~E. 2019, \apjs, 241,
	25
	
	\bibitem[{Ben{\'i}tez-Llambay \& Masset(2016)}]{Benitez-Llambay2016}
	Ben{\'i}tez-Llambay, P., \& Masset, F.~S. 2016, \apjs, 223, 11
	
	\bibitem[{{Ben{\'{\i}}tez-Llambay} \& {Pessah}(2018)}]{Benitez-Llambay2018}
	{Ben{\'{\i}}tez-Llambay}, P., \& {Pessah}, M.~E. 2018, \apjl, 855, L28
	
	\bibitem[{{B{\'e}thune} {et~al.}(2016){B{\'e}thune}, {Lesur}, \&
		{Ferreira}}]{Bethune2016}
	{B{\'e}thune}, W., {Lesur}, G., \& {Ferreira}, J. 2016, \aap, 589, A87
	
	\bibitem[{{Chen} \& {Lin}(2018)}]{Chen2018}
	{Chen}, J.-W., \& {Lin}, M.-K. 2018, \mnras, 478, 2737
	
	\bibitem[{{Dittrich} {et~al.}(2013){Dittrich}, {Klahr}, \&
		{Johansen}}]{Dittrich2013}
	{Dittrich}, K., {Klahr}, H., \& {Johansen}, A. 2013, \apj, 763, 117
	
	\bibitem[{{Dohnanyi}(1969)}]{Dohnanyi1969}
	{Dohnanyi}, J.~S. 1969, \jgr, 74, 2531
	
	\bibitem[{{Garaud} {et~al.}(2004){Garaud}, {Barri{\`e}re-Fouchet}, \&
		{Lin}}]{Garaud2004}
	{Garaud}, P., {Barri{\`e}re-Fouchet}, L., \& {Lin}, D.~N.~C. 2004, \apj, 603,
	292
	
	\bibitem[{Hunter(2007)}]{Hunter2007}
	Hunter, J.~D. 2007, Computing In Science \& Engineering, 9, 90
	
	\bibitem[{Jacquet {et~al.}(2011)Jacquet, Balbus, \& Latter}]{Jacquet2011}
	Jacquet, E., Balbus, S., \& Latter, H. 2011, \mnras, 415, 3591
	
	\bibitem[{{Johansen} \& {Youdin}(2007)}]{Johansen2007}
	{Johansen}, A., \& {Youdin}, A. 2007, \apj, 662, 627
	
	\bibitem[{{Johansen} {et~al.}(2009){Johansen}, {Youdin}, \&
		{Klahr}}]{Johansen2009}
	{Johansen}, A., {Youdin}, A., \& {Klahr}, H. 2009, \apj, 697, 1269
	
	\bibitem[{Kowalik {et~al.}(2013)Kowalik, Hanasz, W{\'o}lta{\'n}ski, \&
		Gawryszczak}]{Kowalik2013a}
	Kowalik, K., Hanasz, M., W{\'o}lta{\'n}ski, D., \& Gawryszczak, A. 2013,
	\mnras, 434, 1460
	
	\bibitem[{{Krapp} {et~al.}(2018){Krapp}, {Gressel}, {Ben{\'{\i}}tez-Llambay},
		{Downes}, {Mohandas}, \& {Pessah}}]{Krapp2018}
	{Krapp}, L., {Gressel}, O., {Ben{\'{\i}}tez-Llambay}, P., {et~al.} 2018, \apj,
	865, 105
	
	\bibitem[{{Laibe} \& {Price}(2014)}]{Laibe2014}
	{Laibe}, G., \& {Price}, D.~J. 2014, \mnras, 440, 2136
	
	\bibitem[{{Lin}(2019)}]{Lin2019}
	{Lin}, M.-K. 2019, \mnras, 485, 5221
	
	\bibitem[{{Lin} \& {Youdin}(2017)}]{Lin2017}
	{Lin}, M.-K., \& {Youdin}, A.~N. 2017, \apj, 849, 129
	
	\bibitem[{{Mathis} {et~al.}(1977){Mathis}, {Rumpl}, \&
		{Nordsieck}}]{Mathis1977}
	{Mathis}, J.~S., {Rumpl}, W., \& {Nordsieck}, K.~H. 1977, \apj, 217, 425
	
	\bibitem[{{Nakagawa} {et~al.}(1986){Nakagawa}, {Sekiya}, \&
		{Hayashi}}]{Nakagawa1986}
	{Nakagawa}, Y., {Sekiya}, M., \& {Hayashi}, C. 1986, \icarus, 67, 375
	
	\bibitem[{P\'erez \& Granger(2007)}]{Perez2007}
	P\'erez, F., \& Granger, B.~E. 2007, Computing in Science and Engineering, 9,
	21
	
	\bibitem[{{Pinilla} {et~al.}(2012){Pinilla}, {Birnstiel}, {Ricci}, {Dullemond},
		{Uribe}, {Testi}, \& {Natta}}]{Pinilla2012}
	{Pinilla}, P., {Birnstiel}, T., {Ricci}, L., {et~al.} 2012, \aap, 538, A114
	
	\bibitem[{{Raettig} {et~al.}(2015){Raettig}, {Klahr}, \& {Lyra}}]{Raettig2015}
	{Raettig}, N., {Klahr}, H., \& {Lyra}, W. 2015, \apj, 804, 35
	
	\bibitem[{{Ragusa} {et~al.}(2017){Ragusa}, {Dipierro}, {Lodato}, {Laibe}, \&
		{Price}}]{Ragusa2017}
	{Ragusa}, E., {Dipierro}, G., {Lodato}, G., {Laibe}, G., \& {Price}, D.~J.
	2017, \mnras, 464, 1449
	
	\bibitem[{{Riols} \& {Lesur}(2018)}]{Riols2018}
	{Riols}, A., \& {Lesur}, G. 2018, \aap, 617, A117
	
	\bibitem[{{Schaffer} {et~al.}(2018){Schaffer}, {Yang}, \&
		{Johansen}}]{Schaffer2018}
	{Schaffer}, N., {Yang}, C.-C., \& {Johansen}, A. 2018, \aap, 618, A75
	
	\bibitem[{{Squire} \& {Hopkins}(2018)}]{Squire2018}
	{Squire}, J., \& {Hopkins}, P.~F. 2018, \mnras, 477, 5011
	
	\bibitem[{Testi {et~al.}(2014)Testi, Birnstiel, Ricci, Andrews, Blum,
		Carpenter, Dominik, Isella, Natta, Williams, \& Wilner}]{testi2014}
	Testi, L., Birnstiel, T., Ricci, L., {et~al.} 2014, Protostars and Planets VI,
	339
	
	\bibitem[{Walt {et~al.}(2011)Walt, Colbert, \& Varoquaux}]{Walt2011}
	Walt, S. v.~d., Colbert, S.~C., \& Varoquaux, G. 2011, Computing in Science and
	Engg., 13, 22
	
	\bibitem[{Weber {et~al.}(2018)Weber, Ben{\'{i}}tez-Llambay, Gressel, Krapp, \&
		Pessah}]{Weber2018}
	Weber, P., Ben{\'{i}}tez-Llambay, P., Gressel, O., Krapp, L., \& Pessah, M.
	2018, \apj, 854, 153
	
	\bibitem[{Weidenschilling(1977)}]{Weidenschilling1977}
	Weidenschilling, S.~J. 1977, \mnras, 180, 57
	
	\bibitem[{{Whipple}(1972)}]{Whipple1972}
	{Whipple}, F.~L. 1972, in From Plasma to Planet, ed. A.~{Elvius}, 211
	
	\bibitem[{Youdin \& Johansen(2007)}]{Youdin2007}
	Youdin, A., \& Johansen, A. 2007, \apj, 662, 613
	
	\bibitem[{Youdin \& Goodman(2005)}]{Youdin2005}
	Youdin, A.~N., \& Goodman, J. 2005, \apj, 620, 459
	
	\bibitem[{Zhu {et~al.}(2012)Zhu, Nelson, Dong, Espaillat, \&
		Hartmann}]{Zhu2012}
	Zhu, Z., Nelson, R.~P., Dong, R., Espaillat, C., \& Hartmann, L. 2012, \apj,
	755, 6
	
\end{thebibliography}

\end{document}